\documentclass[titlepage]{article}
\usepackage{graphicx}
\begin{document}

\title{Strong Electromagnetic Waves Propagation in an Electron-Positron-Ion Plasma}
\author{M. Y. EL-Ashry\thanks{
Physics Department, Faculty of Science, Suez Canal Universitry, Ismailia,
EGYPT. } \and E. I. Shalaan\thanks{
The same address as of the first author }}
\maketitle

\begin{abstract}
The main purpose of this work is the investigation of the influence of
electron-positron-ion on the electrostatic wakefields that are deriven by
intense, short electromagnetic wave packets in a three component
unmagnetized plasma. The equations that describes the deriven electrostatic
wakefield are given in the form of system of nonlinear differential
equations. The solution of that system was carried out numerically and the
results are displayed graphically.
\end{abstract}

%
%
%
%
%
%

\section*{Introduction}

Recently, the nonlinear propagation of electromagnetic waves in
electron-positron plasma has attracted the interest of researchers\cite
{shukla} \cite{t2} and\cite{vib}. Due to the fact that the electron-positron
plasma are found in Van Allen Belts, near polar cap of pulsar, in the active
galactic nuclei, as well as in the early universe.

When plasma becomes so hot that it becomes relativistic the temperature $T $
of the plasma exceeds the rest mass energy of electrons $m c^2 = 0.5 $ MeV.
In this relativistic regime the processes of electron-positron pair creation
and annihilation become important; $2 \gamma \rightleftharpoons e^+ + e^- $.
In relativistic temperature the electrons (and positrons) energy $E_e $ far
exceeds the rest mass energy so that electrons and positrons behave
kinematically similar to photons and come into equilibrium with nearly equal
population.

In 1990 T. Tajima and T. Taniuti \cite{tji} are interested in the process of
electron-positron plasma creation and annihilation occurs in relativistic;
unmagnetized plasma at high temperature of the plasma exceeds the rest mass
energy of the electrons. They are emphasize the dominant population of
electrons and positrons and neglect ions effect.

In \cite{surko2}, it was shown that , the positrons can be used to probe the
particle transport in Tokamak plasma. This new diagnostic technique could
also be useful in studying transport in other magnetic confinement devices
such as reversed-field pinches and magnetic mirrors. Electron-positron pair
production can also be possible during intense short laser pulse propagation
in plasma. When picosecond electromagnetic pulse having intensity beyond $%
10^{21}$ W/$\mbox{cm}^2$ interact with plasma, the plasma electrons can
acquire oscillatory velocity ($V_{os}=eE_o/mc$) higher than the speed of
light $c$. Here $E_o(W_o)$ is the amplitude (frequency) of the $EM$
radiation, $c$ \& $m$ are respectively the magnitude of the electron charge
and the rest mass of the electron.

Several authors have suggested that when the oscillatory electron energy $%
E_{os} = m c^2 ( 1 + \frac{V^2_{os}}{c^2})^{1/2} $ exceeds $3 m c^2 $ then
these electrons can create an electron- positron pairs in the presence of
background positive ions. And also it has been shown that powerful short
laser pulsar can generate a large amplitude electrostatic wave (Wakefield)
in a plasma in which the electron plasma frequency $\omega_{pe} $ is much
smaller than $\omega_o $. Since the amplitude of the excited wakefield can
be very high ( $e \phi / m_o c^2 \equiv \Phi \gg 1 $, where $\phi $ is the
electrostatic potential ), the plasma electrons also acquire the
longitudinal energy $E_L = m_o c^2 ( 1 + P^2_\parallel /( m_o c^2)) \gg 3
m_o c^2$, where $P_\parallel $ is the relativistic momentum in the
wakefield. Thus, there is the possibility of creating electron-positron
pairs in those regions in which wakefields and short $EM $ pulses are
localized. Since the lifetime $\tau_p > \omega_p^{-1} $ of the positron in
such a plasma is sufficiently long.

We shall have a plasma that is an admixture of electrons, positrons and
positive ions. A three-component electrons-positron-ion plasma can indeed be
created in laboratory plasma. V. I. Berezhiani, and et. al. \cite{vib} have
investigated the influence of electron-positron pairs on the electrostatic
wakefield that are driven by intense, short $EM$ wave packets in a three
component unmagnetized plasma. They found that in contrast to pure
electron-ion plasma, the presence of positrons significantly reduce both the
amplitude and the wavelength of wakefields. In view of their investigation,
it is suggestive that the production of electron-positron pairs by
wakefields in Laboratory and cosmic plasma ought to be reconsidered because
the presence of pairs may affect the wakefield that is responsible for
accelerating electrons to very high energy.

The important of three component admixture plasma had led to much related
theoretical investigations. Of these, Rizzato \cite{riz}, investigated the
possibility of high-frequency $EM$ localization in cold unmagnetized plasmas
made up of electrons, positrons and ions. It was shown that such a
possibility depends on the concentration of particles, on the velocity of
the solitons and on the angle between the direction of modulation and the
direction of the fast spatial dependence. In particular in the case of pure
electron-positron plasmas Rizzato was shown that no localized solution is
possible which is in contrast with results derived by Mofiz et. al. (1985)
from a similar model. Also it was shown that low frequency magnetic fields
may appear in the case of oblique modulation and that both the amplitude of
the magnetic field and the amplitude of the solitary wave are very sensitive
functions of the angle between the direction of modulation and fast spatial
dependence.

Berezhiani, et. al. \cite{shukla} analyzed the nonlinear interaction of an
arbitrarily large amplitude circularly polarized $EM$ wave with an
unmagnetized electron-positron plasma. taking into account relativistic
particle-mass variation as well as large-scale density perturbation created
by radiation pressure. It was found that the interaction is governed by an
equation for the electromagnetic wave envelope, which is coupled with a
pairs of equations describing fully nonlinear longitudinal plasma motion.
This results should be useful for the understanding of nonlinear photon
motion in cosmic plasmas, such as those found in the early universe and
active galactic nuclei.

Mofiz \cite{mofiz2} studied the nonlinear propagation of Langmiur wave in a
hot ultrarelativistic electron-positron plasma. It was shown that, in a
dense ultrarelativistic electron-positron plasma electrostatic modes with
wavelength greater than the plasma Debye length produce Langmiur Solitons
that are spiky in nature. One essential feature of these solitons is that
they can not form an energy flow toward smaller size, since they can not
merge with each other. To merge, the soliton would have to give energy to
sound waves, which do not exist in electron-positron plasma. In the one
dimensional case the only process accruing in a soliton gas is nonlinear
interaction with electron and positron, i. e. nonlinear Landau damping,
which stops the solitons without changing its amplitude. The stopping length
is of order of the soliton width, which is very small for a dense plasma.
Thus the soliton will be stopped in a short time, of the order of their
creation. these spiky short duration Langmiur solitons might berelated to
the pulsar radiation and with its microstructures.

In \cite{ash}, V.I.Berezhiani, M.Y. El-Ashry and U.A. Mofiz are investigated
the propagation of intense electromagnetic radiation in an admixture of
unmagnetized electron-positron-ion plasma analytically. It was shown that
electromagnetic radiation of arbitrary amplitude in presence of heavy ions,
in contrast to the case of pure electron-positron plasma, may be localized
with the generation of humped ambipolar potential in the plasma, i.e. the
driving field intensity creates intense soliton in the plasma with the
generation of double hump ambipolar potentials. With the increase of the
value of $\epsilon$ (=$n_{op}/n_{oe}$), they found that the tendency of
converging a single hump soliton to a double hump one. In their
investigation they neglected the ion dynamics. Consideration of ion
dynamics, of course, may effect the localization phenomenon. This effect
will be studied here.

Now, the vector potential {\bf A} for circularly polarized {\bf EM} wave can
be expressed as:
\[
{\bf A}={\bf A_{\perp }}+{\bf A_z}
\]
and
\begin{eqnarray*}
{\bf A_{\perp }} &=&{\bf a}(z,t)\;e^{\textstyle i(kz-\omega t)}+c.c. \\
{\bf A_z} &=&{\bf \acute{a}}(z,t)\;e^{\textstyle i(kz-\omega t)}+c.c.
\end{eqnarray*}
where:
\begin{eqnarray*}
{\bf a}(z,t) &=&a(z-Vt)({\bf \hat{x}}+i{\bf \hat{y}}) \\
{\bf \acute{a}}(z,t) &=&\acute{a}(z-Vt){\bf \hat{z}}
\end{eqnarray*}
and $a$ and $\acute{a}$ are real functions.

To describe the admixture of plasma made up of electrons, positrons and
ions, we use Maxwell equations, in which the fields are expressed in terms
of the potentials, as described above. Accordingly, using Poissons equation,
we may obtain the following equations for the potentials:
\begin{equation}
\frac{\partial ^2}{\partial t^2}{\bf A_{\perp }}-c^2\nabla ^2{\bf A_{\perp }}%
=4\pi c{\bf J}-c\frac \partial {\partial t}\nabla \phi   \label{c332}
\end{equation}
and:
\begin{equation}
\nabla ^2\phi =-4\pi \rho   \label{3}
\end{equation}
here, ${\bf \rho }$ and ${\bf J}$ are the charge and current densities, and
given by:
\begin{eqnarray*}
\rho  & = & \sum_\alpha e_\alpha n_\alpha  \\
{\bf J} & = &\sum_\alpha e_\alpha n_\alpha {\bf V_\alpha }
\end{eqnarray*}
where $\alpha $ indicates the particle species $\alpha $ (=e, p, i
for electrons, positrons, and ions, respectively), $e_\alpha $ and
$n_\alpha $ are the charge and density of the corresponding
particle $\alpha $. We shall consider the case in which the
admixture equilibrium state is characterized by
$n_{oe}=n_{op}+n_{oi}$, where $n_{o\alpha }$, is the equilibrium
density of the particle $\alpha $.

The relativistic equations of motion of different particles of unmagnetized
plasma admixture is written as:
\begin{equation}  \label{epi1}
\frac{\partial}{\partial t} {\bf P_\alpha} + m_\alpha c^2 \nabla
\gamma_\alpha = e_\alpha \Big[ - \nabla \phi - \frac{1}{c} \frac{\partial
{\bf A_\perp}}{\partial t} \Big]
\end{equation}
where
\[
{\bf P_\alpha} = m_\alpha \gamma_\alpha {\bf V_\alpha}
\]
and
\[
\gamma_\alpha = \bigg( 1 - \frac{v^2_\alpha}{c^2} \bigg)^{-1/2} = \bigg( 1 +
\frac{P^2_\alpha}{m^2_\alpha c^2} \bigg)^{1/2}
\]
where, $m_\alpha$ is the rest mass of the particle $\alpha$.

The continuity equation for the particle $\alpha$ is:
\begin{equation}  \label{7}
\frac{\partial n_\alpha}{\partial t} + \nabla \cdot (n_\alpha {\bf V_\alpha}%
) = 0
\end{equation}

We are looking for localized one-dimensional solution of this system of
equation for a circularly polarized EM wave, where the vector potential {\bf %
A} can be expressed as:
\[
{\bf A} = \frac{1}{2} ({\bf \hat{x} + i \hat{y}}) A(\zeta) e^{\textstyle i
(kz - \omega t)} + c.c.
\]
here, $\zeta = z - V t$.

Analyzing the equation of motion and the wave equations into longitudinal
(parallel to the direction of the wave propagation) and a transverse
(perpendicular plan) parts, then the transverse part of the equation of
motion is immediately integrated to give:
\begin{equation}  \label{epi2}
{\bf P_{\alpha \perp}} = - \frac{e_\alpha}{c} {\bf A_\perp}
\end{equation}
where the constant of integration is set equal to zero, since the particles
were assumed to be immobile at infinity where the field is zero.

Meanwhile, the longitudinal part of the equation of motion takes the
following form:
\[
\frac{\partial P_{\alpha z}}{\partial t} + m_\alpha c^2 \frac{\partial}{%
\partial z}\bigg(1+ \frac{P^2_{\alpha \perp} + P^2_{\alpha z}}{m_\alpha^2 c^2%
}\bigg)^{1/2}
\]
\begin{equation}
= - e_\alpha \frac{\partial \phi}{\partial z}  \label{epi3}
\end{equation}

And, accordingly, equations(~\ref{c332}), and (~\ref{3}) may be rewritten
as:
\[
\bigg( \frac{\partial^2}{\partial z^2} - \frac{1}{c^2} \frac{\partial^2}{%
\partial t^2} \bigg){\bf A_\perp} = - \frac{4 \pi e}{c} (n_p {\bf V_p}
\]
\begin{equation}
+ n_i {\bf V_i} - n_e {\bf V_e})  \label{epi4}
\end{equation}
\begin{equation}  \label{epi5}
\frac{\partial^2 \phi}{\partial z^2} = - 4 \pi e ( \delta n_p + \delta n_i
-\delta n_e)
\end{equation}

Now, we consider only the electron; but everything stated below applies
equally well, of course, to positron and ion.

We have:
\[
\frac{\partial}{\partial t} = \frac{\partial}{\partial \zeta} \frac{\partial
\zeta}{\partial t} = - v_g \frac{\partial}{\partial \zeta}
\]
and also
\[
\frac{\partial}{\partial z} = \frac{\partial}{\partial \zeta} \frac{\partial
\zeta}{\partial z} = \frac{\partial}{\partial \zeta}
\]
then, equation(~\ref{epi3}) reduced to
\[
-v_g \frac{\partial}{\partial \zeta} P_{ez} + m c^2 \frac{\partial}{\partial
\zeta} \sqrt{1 + \frac{P^2_{ez}+P^2_{e \perp}}{m^2 c^2}} = e \frac{\partial
\phi}{\partial \zeta}
\]

Now, we can integrate both sides with respect to $\zeta$, yield:
\[
-v_g P_{ez} + m c^2 \sqrt{1 + \frac{P^2_{ez}+P^2_{e \perp}}{m^2 c^2}} = e
\phi + C
\]
where, $C$ is the constant of integration, which could be determine from the
boundary conditions. Here, $P_\perp,P_{ez} \rightarrow 0 \mbox{and} \phi
\rightarrow 0 $ when $\zeta \rightarrow 0 $, we get; $C=mc^2$, and
accordingly:
\[
- \left(\frac{v_g}{c}\right)\left(\frac{P_{ez}}{m c }\right) + \sqrt{1 +
\frac{P^2_{ez}}{m^2 c^2}+ \frac{P^2_{e \perp}}{m^2 c^2}}
\]
\begin{equation}  \label{epi6}
= \frac{e \phi}{m c^2} + 1
\end{equation}

Let:
\begin{eqnarray*}
\Gamma_{\alpha z} & = & \frac{P_{\alpha z}}{m c^2} \\
U_g & = & \frac{v_g}{c} \\
\Psi_\alpha & =&- \frac{P_{\perp \alpha}}{m_\alpha c^2} \\
\Phi_\alpha & = &- \frac{e_\alpha \phi}{m_\alpha c^2}
\end{eqnarray*}
but for simplicity we will drop here the subscript e; i.e. $\Phi_e = \Phi %
\mbox{and} \Psi_e = \Psi $

Equation(~\ref{epi6}) becomes
\[
- U_g \Gamma_{e z} + \sqrt{1 + \Gamma^2_{e z} + \Psi^2 } = \Phi + 1
\]

We note here that the began value of $\Gamma_{ez}$ is considered to be zero,
then the above eaquation yields:
\[
\Gamma_{ez} = U_g (\Phi + 1) \gamma_g^2 -
\]
\begin{equation}  \label{epi9}
\gamma^2_g \sqrt{ (\Phi + 1)^2 - \frac{1 + \Psi^2}{\gamma_g^2}}
\end{equation}
where $\gamma_g = \frac{1}{\sqrt{1 - U_g^2}}$

If we return back to the continuity equation; we get:
\[
-v_g \frac{\partial n_e}{\partial \zeta} + n_e \frac{\partial v_{ez}}{%
\partial \zeta} =0
\]
by integrating:
\[
-v_g n_e + n_e v_{ez} = C
\]
here, $v_{ez} \rightarrow 0$ when $n_e \rightarrow n_{oe}$, then, $C=- v_g
n_{oe} $, and accordingly:
\[
n_e = \frac{U_g n_{oe} \sqrt{1+\Gamma_{ez}+\Psi^2}}{U_g \sqrt{%
1+\Gamma_{ez}+\Psi^2} - \Gamma_{ez}}
\]
but; $\delta n_e = n_e - n_{oe}$, hence
\begin{equation}
\delta n_e = \frac{ n_{oe} \Gamma_{ez}}{U_g \sqrt{1+\Gamma_{ez}+\Psi^2} -
\Gamma_{ez}}  \label{epi12}
\end{equation}

From equations(~\ref{epi9}) and (~\ref{epi12}), we get:
\begin{equation}  \label{epi13}
\delta n_e = n_{oe} \gamma^2_g \left(\frac{U_g (1+\Phi ) }{\sqrt{(1+\Phi)^2
- \frac{1 + \Psi^2}{\gamma^2_g}} }-1\right)
\end{equation}
here, we should note that:
\begin{eqnarray*}
\Phi_e &=& \Phi \\
\Phi_p &=& -\Phi \\
\Phi_i &=& - \frac{m}{M} \Phi
\end{eqnarray*}
where $M$ is the mass of an ion. and that:
\begin{eqnarray*}
\Psi_e &=& \Psi \\
\Psi_p &=& -\Psi \\
\Psi_i &=& - \frac{m}{M} \Psi
\end{eqnarray*}

By follows similar steps; for positron and ion; we can get the following:
\begin{equation}  \label{epi14}
\delta n_p = n_{op} \gamma^2_g \left( \frac{ U_g (1-\Phi ) }{\sqrt{%
(1-\Phi)^2 - \frac{1 + \Psi^2}{\gamma^2_g}} }-1 \right)
\end{equation}
and
\begin{equation}  \label{epi15}
\delta n_i = n_{oi} \gamma^2_g \left( \frac{ U_g (1- \frac{m}{M}\Phi ) }{%
\sqrt{(1-\frac{m}{M}\Phi)^2 - \frac{1 + \frac{m}{M}\Psi^2}{\gamma^2_g}} }-1
\right)
\end{equation}

The equation(~\ref{epi5}) can be rewritten as:
\begin{equation}  \label{epi16}
\frac{\partial^2}{\partial z^2} \Phi = \frac{4 \pi e^2}{m c^2} (\delta n_e -
\delta n_p - \delta n_i)
\end{equation}

Substituting equations(~\ref{epi13}), (~\ref{epi14}) and (~\ref{epi15}) in
equation(~\ref{epi16}):
\begin{eqnarray*}
\frac{\partial^2}{\partial z^2} \Phi &=& \left( \frac{\omega^2}{c^2}\right)
\gamma^2_g U_g \Bigg[ \left( \frac{ U_g (1+\Phi ) }{\sqrt{(1+\Phi)^2 - \frac{%
1 + \Psi^2}{\gamma^2_g}} } \right) \\
&&- \epsilon_1 \left( \frac{ U_g (1-\Phi ) }{\sqrt{(1-\Phi)^2 - \frac{1 +
\Psi^2}{\gamma^2_g}} } \right) \\
&&- \epsilon_2 \left( \frac{ U_g (1-\frac{m}{M}\Phi ) }{\sqrt{(1-\frac{m}{M}%
\Phi)^2 - \frac{1 +\frac{m}{M} \Psi^2}{\gamma^2_g}} } \right) \\
&&- - \frac{1-\epsilon_1 - \epsilon_2}{U_g} \Bigg]
\end{eqnarray*}

But; $1-\epsilon_1 - \epsilon_2 = 1 - \frac{n_{op}}{n_{oe}} - \frac{n_{oi}}{%
n_{oe}}= \frac{n_{oe} - n_{op} - n_{oi}}{n_{oe}} $ and $n_{oe}=n_{op}+n_{oi}
$; we get, $1-\epsilon_1 - \epsilon_2 = 0$, hence:
\begin{eqnarray*}
\frac{\partial^2}{\partial z^2} \Phi &=& \left( \frac{\omega^2}{c^2}\right)
\gamma^2_g U_g \Bigg( \frac{ U_g (1+\Phi ) }{\sqrt{(1+\Phi)^2 - \frac{1 +
\Psi^2}{\gamma^2_g}} } \\
&&- \epsilon_1 \frac{ U_g (1-\Phi ) }{\sqrt{(1-\Phi)^2 - \frac{1 + \Psi^2}{%
\gamma^2_g}} } \\
&&- \epsilon_2 \frac{ U_g (1-\frac{m}{M}\Phi ) }{\sqrt{(1-\frac{m}{M}\Phi)^2
- \frac{1 +\frac{m}{M} \Psi^2}{\gamma^2_g}} } \Bigg)
\end{eqnarray*}

If we let $\frac{\partial^2}{\partial z^2} = \frac{\partial^2}{\partial
\zeta^2}= \frac{\omega^2}{c^2} \frac{\partial^2}{\partial \eta^2} $, we can
write:
\begin{eqnarray}
\frac{\partial^2 \Phi}{\partial \eta^2} &=& \gamma^2_g U_g^2 \Bigg( \frac{
(1+\Phi ) }{\sqrt{(1+\Phi)^2 - \frac{1 + \Psi^2}{\gamma^2_g}} }  \nonumber \\
&&- \frac{\epsilon_1 (1-\Phi ) }{\sqrt{(1-\Phi)^2 - \frac{1 + \Psi^2}{%
\gamma^2_g}} }  \nonumber \\
&&- \frac{\epsilon_2 (1-\frac{m}{M}\Phi ) }{\sqrt{(1-\frac{m}{M}\Phi)^2 -
\frac{1 +\frac{m}{M} \Psi^2}{\gamma^2_g}} } \Bigg)  \label{phi}
\end{eqnarray}

Equation(~\ref{epi4}) can be written as:
\[
\bigg( \frac{\partial^2}{\partial z^2} - \frac{1}{c^2} \frac{\partial^2}{%
\partial t^2} \bigg){\bf A_\perp} =
\]
\[
\frac{4 \pi e^2}{m c^2} \left(\frac{n_e }{\gamma_e} + \frac{n_p }{\gamma_p}
+ \frac{m n_i }{M \gamma_i}\right) A_\perp
\]

By follows similar steps; the last equation can be written as:
\begin{eqnarray}
\frac{\partial^2 \Psi}{\partial \eta^2} &=& U_g \gamma^2_g \Psi \Bigg(\frac{1%
}{\sqrt{ (1 + \Phi )^2 - \frac{1 + \Psi^2}{\gamma_g^2}}}  \nonumber \\
&& +\frac{\epsilon_1}{\sqrt{ (1 - \Phi )^2 - \frac{1 + \Psi^2}{\gamma_g^2}}}
\nonumber \\
&&+ \frac{(m/M) \epsilon_2}{\sqrt{ (1 - \frac{m}{M} \Phi )^2 - \frac{1 +
\frac{m}{M} \Psi^2}{\gamma_g^2}}} \Bigg)  \label{psi}
\end{eqnarray}

Equation(~\ref{phi}) and equation(~\ref{psi}) are coupled together and
forming a system of nonlinear second order deferential equations. It is
rather difficult to obtain an analytical solution of them. However, in the
following, we present numerical solutions of equation(~\ref{phi}) and
equation(~\ref{psi}). Taking some typical parameters:
\begin{eqnarray*}
\frac{1}{\gamma^2_g} &= & 0.12 \\
\frac{m}{M} & = & \frac{1}{1834} \\
U_g & =& \sqrt{1 - 0.12}
\end{eqnarray*}
and by using Taylor method, we have obtained the wakefield profiles at
different values of $\epsilon_1$ and $\epsilon_2$. The results are displayed
in the following figures:

\pagebreak

\begin{center}
\includegraphics{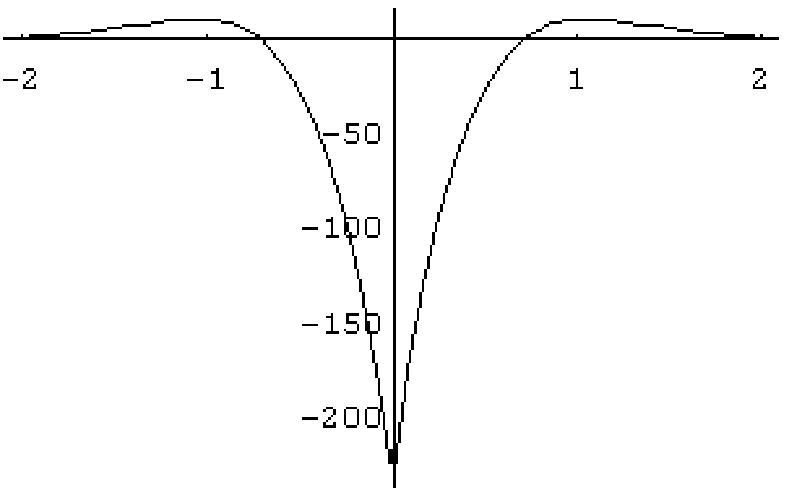}
\end{center}
\center{Fig1-a: The wakefield profiles for $\Psi $ at $\epsilon
_1=0.3,\epsilon _2=0.7$ \vspace{1.5cm} }

\vspace{1.5cm}

\begin{center}
\includegraphics{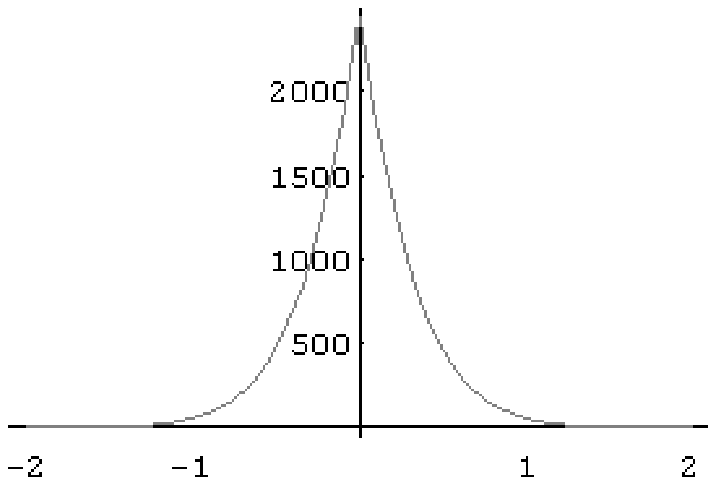}
\end{center}
\center{{Fig1-b: The wakefield profiles for $\Phi $ at $\epsilon
_1=0.3,\epsilon _2=0.7$} \vspace{1.5cm}}

\pagebreak

\begin{center}
\includegraphics{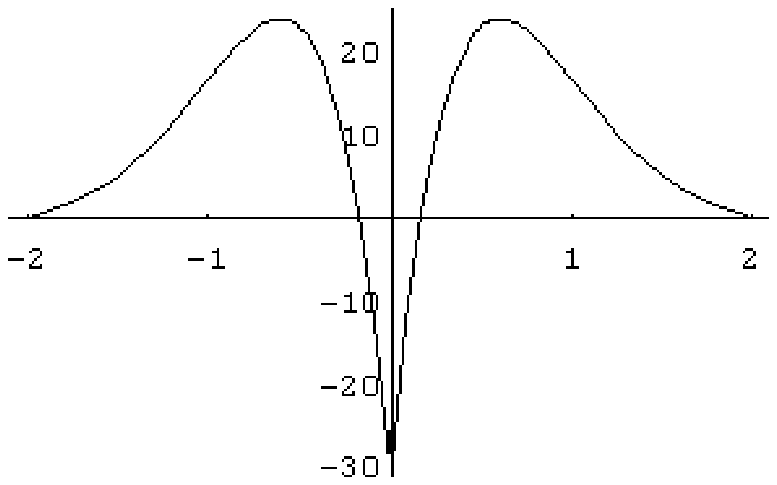}
\end{center}
\center{ Fig2-a: The wakefield profiles for $\Phi $ at $\epsilon
_1=0.4,\epsilon _2=0.6$}

\vspace{1.5cm}

\begin{center}
\includegraphics{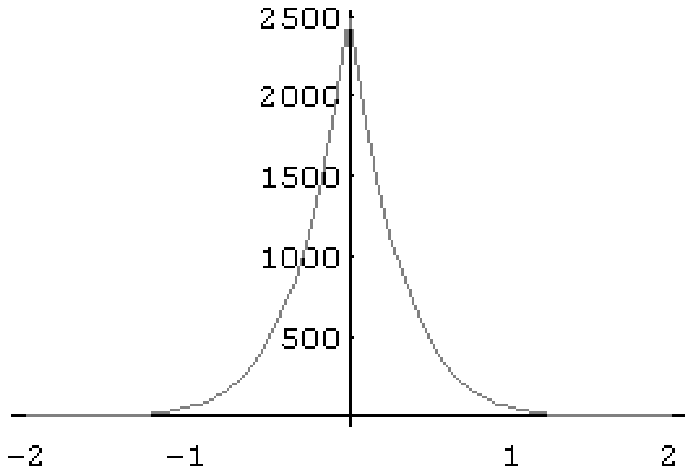}
\end{center}
\center{ Fig2-b: The wakefield profiles for $\Phi $ at $\epsilon
_1=0.4,\epsilon _2=0.6$ }

\pagebreak


\end{document}